\documentclass{article}
\usepackage{spconf,amsmath,graphicx}
\usepackage{float} 
\pdfoutput=1
\usepackage{verbatim}
\usepackage{subfig}
\usepackage{cite}
\usepackage{stfloats}
\usepackage{hyperref}
\usepackage{subcaption}
\usepackage{booktabs}
\usepackage{multirow}
\usepackage{caption}

\title{EMO-Codec:  An In-Depth Look at Emotion Preservation capacity of Legacy and Neural Codec Models With Subjective and Objective Evaluations}
\name{\parbox{0.8\linewidth}{\centering 
    Wenze Ren$^{1*}$, Yi-Cheng Lin$^{1*}$, Huang-Cheng Chou$^2$, Haibin Wu$^1$, Yi-Chiao Wu, Chi-Chun Lee$^2$, Hung-yi Lee$^1$, Yu Tsao$^3$}
}

\address{National Taiwan University$^1$, National Tsinghua University$^2$, Academia Sinica$^3$}

\begin{document}
\ninept
\maketitle
\def\thefootnote{*}\footnotetext{Equal first contribution}
\begin{abstract}

The neural codec model reduces speech data transmission delay and serves as the foundational tokenizer for speech language models (speech LMs). Preserving emotional information in codecs is crucial for effective communication and context understanding. However, there is a lack of studies on emotion loss in existing codecs. This paper evaluates neural and legacy codecs using subjective and objective methods on emotion datasets like IEMOCAP. Our study identifies which codecs best preserve emotional information under various bitrate scenarios. We found that training codec models with both English and Chinese data had limited success in retaining emotional information in Chinese.
Additionally, resynthesizing speech through these codecs degrades the performance of speech emotion recognition (SER), particularly for emotions like sadness, depression, fear, and disgust. Human listening tests confirmed these findings. This work guides future speech technology developments to ensure new codecs maintain the integrity of emotional information in speech.

\end{abstract}

\section{Introduction}
\label{sec:intro}

Audio codecs are first introduced to compress speech signals to a limited number of bits for low-latency communication. 
They typically include both an encoder and a decoder to reconstruct the compressed codes back into speech signals. Given their success, researchers are exploring how the advanced capacity of LMs, which have demonstrated remarkable performance in text modeling and surpassing human abilities across various tasks, can integrate speech perception capacity.\cite{touvron2023llama, anil2023palm, alpaca}. 
Recent advancements in speech LMs use codec codes as discrete tokens. \cite{anygpt, speechx, viola, vallex, valle, uniaudio, borsos2023soundstorm, wang2024lauragpt, kuan2023towards}
What distinguishes speech LMs is their capacity to extract richer information, such as emotion, from spoken language \cite{huang2023dynamicsuperb, bai2023qwen}.
They capture content information and delve into the nuances of speaker identity and emotion, aspects that text alone cannot fully grasp.

Due to the wide application of codec models for reducing communication latency and serving as tokenizers for speech LMs, the codecs should preserve the signal's integrity and substantial emotional information. 
For example, when an individual communicates with a virtual assistant through voice commands, the emotional information embedded within the speech signal can provide valuable context for the assistant to deliver more empathetic and tailored responses \cite{guha2022desco}.
However, the speech codec used in the communication pipeline inadvertently distorts or discards critical emotional cues during compression. 
In that case, the assistant may struggle to perceive the user's emotional state accurately, leading to less effective interactions. 
Therefore, preserving emotion in speech signals is essential for the effectiveness of speech LMs.

Many advanced neural codec models have been developed using different techniques \cite{wu2024audio}. 
However, existing evaluations on these codecs predominantly focus on signal-level metrics, overlooking crucial paralinguistic elements like emotion \cite{9414901}. 
While Codec-SUPERB \cite{wu2024codecsuperb} endeavors to compare emotion preservation in neural codec reconstructed speech, its assessment is confined to a single small dataset, single language, and a specific case-study model.
Similarly, Siegert et al. \cite{7776178} investigate the intelligibility of codec-compressed emotional speech but overlook the evaluation of affective content in speech. 
Previous works compared the speech distorted by legacy codec compression algorithms and evaluated speech emotion recognition (SER) performance by human perception \cite{niebuhr2022high} and automatic methods such as Gaussian mixture model \cite{7843353, 7330399, albahri2016effect} or Support Vector Machine (SVM) \cite{Lotz2017_212}.
However, these studies are confined to legacy codecs, which exhibit inferior performance compared to neural codecs.
They also utilize evaluation models that are less precise than advanced SER models.


\begin{figure*}[!t]
    \centering
    \vspace{-0.5cm}
    \includegraphics[width=\textwidth]{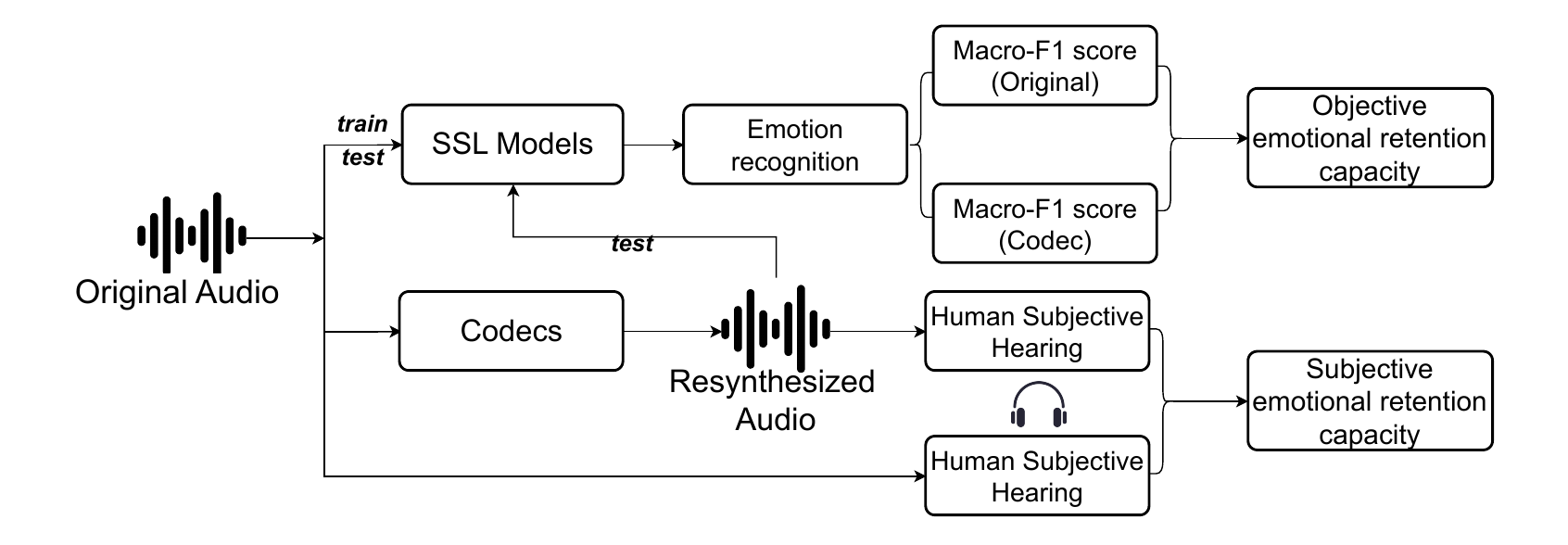}
    \vspace{-0.1cm}
    \caption{The pipeline of Emo-Codec. The process starts by training the SSL model for emotion recognition using only the original audio. Then, we inference the testset of the original audio, and the audio resynthesized with codecs. We calculate the $F_1$ score difference of emotion recognition to get the objective emotion loss. We used the original audio files and the resynthesized audio files from the codec model to do the human subjective listening test for emotion recognition.}
    \label{fig:main_pipeline}
\end{figure*}

There is an urgent need for a comprehensive analysis to compare the ability of codec models in different languages to maintain sentiment across different SER systems under staged or real-world, multilingual, and multi-speaker dataset conditions. We consider various factors of codecs, such as bitrates, pretraining dataset languages, and architecture. Different evaluations are conducted to comprehensively assess these factors and their effects on the accuracy of the pre-trained SER systems and the emotional perception of humans. Our goal is to provide a path for the community to refer for a new design of codecs. Our study comprehensively evaluates the efficacy of 14 neural network codecs and 3 legacy codecs in retaining emotional information across 15 different SER models on six datasets, revealing their potential to enhance affective computing in real-world applications. The main flow of Emo-Codec is shown in the Fig.~\ref{fig:main_pipeline}.


In conclusion, our work contributes the following valuable findings:
\begin{itemize}
    \item Emo-Codec provides comprehensive performance benchmarks for fourteen codec models in six emotion datasets, highlighting their capacity to preserve emotional information.
    \item Descript Audio Codec (DAC) series \cite{kumar2024high} consistently outperforms other codecs in SER at equivalent bit rates. Additionally, AcademiCodec \cite{yang2023hifi} and SpeechTokenizer \cite{zhang2023speechtokenizer} show considerable performance under low-bit rate scenarios.
    \item We empirically find that training codec models using both Chinese and English data provides limited improvements in preserving Chinese emotional information compared with codec models trained with only English data.
    \item Of all the emotions investigated, some negative valence emotions, sadness, depression, fear, and disgust, exhibit a higher performance drop than other emotions.
    
\end{itemize}

\section{Method}
This section first overviews our rationales for the large-scale evaluation and then details each part.

\begin{table}[b]
\centering
\fontsize{7}{9}\selectfont
\caption{Overview of the used SSL models and the number of their parameters (Parm (M)).}
\vspace{-0.3cm}
\begin{tabular}{l|c|l|c}

\toprule
\multicolumn{1}{c|}{\textbf{Model}} & \textbf{Parm (M)} & \multicolumn{1}{c|}{\textbf{Model}} & \textbf{Parm (M)}\\ 
\midrule
Wav2Vec 2.0 XLS-R-1B & 965 & VQ-Wav2Vec & 34\\
WavLM large & 317 & Wav2vec-large-960h & 33\\
Wav2Vec 2.0 Large Robust & 317 & TERA & 21 \\ 
HuBERT Large & 317 & NPC & 19 \\
Wav2Vec2 large 960h & 317 & VQ-APC & 5 \\ 
Data2vec large 960h & 313 & APC & 4 \\ 
DeCoAR 2.0 & 90 & Modified CPC & 2 \\ 
Mockingjay & 85 &  \\ 
\bottomrule
\end{tabular}
\label{tab:model-inform}
\vspace{-0.3cm}
\end{table}

\subsection{Rationales of our evaluation} 
We evaluate both neural and legacy codecs to ensure a comprehensive analysis. The legacy codecs serve as established benchmarks, enabling us to gauge the advancements made by neural codecs.
For neural codecs, we select models based on their functionality and design principles. These codecs encompass a range of innovative architectural designs and sophisticated methodologies, ensuring optimal performance and versatility in processing various types of speech data. Specifically, we target models explicitly tailored for speech language model (LM) tokenization. Additionally, we consider models trained on mixed Chinese and English data such as Funcodec, to ensure comprehensive linguistic coverage, accounting for the nuances and emotional variances inherent in both languages.
We evaluate all codec models at similar bitrates to ensure a fair comparison.\\
For SER models, regarding Table~\ref{tab:model-inform}, we use the representation from self-supervised learning (SSL) models to train SER models because the SSL paradigm has reached state-of-the-art performance on SER tasks \cite{10089511, 9747870}. We adopt various emotion datasets to increase the diversity of language, dataset collection methods (real world, improvised act, scripted act), and speaker.

\begin{table}[t]
\centering
\fontsize{7}{9}\selectfont
\caption{The table summarizes the selected codecs and their transmission bit rate. All codecs in this article are represented by the code name here. The column \textbf{ID} shows the simplified name for all codecs,  \textbf{kBPS} represents the bit rate in kilo-Bit Per Second, and \textbf{sr} shows the sample rate of the codec model.\label{tab:codec-inform}}
\vspace{-0.3cm}
\begin{tabular}{c|c|c|c|c}
\hline
\textbf{ID} & \textbf{Codec} & \textbf{Codec Configuration} & \textbf{kBPS} & \textbf{sr} \\ \hline
U & AudioDec& symAD\_libritts\_24000\_hop300 & 6.4 & 24 \\ \hline
C & AcademiCodec & large universal & 2 & 16\\ \hline
S & SpeechTokenizer & hubert\_avg & 4 & 16\\ \hline
D1 & \multirow{2}{*}{DAC} & DAC\_16k & 6 & 16\\ 
D2 & & DAC\_24k & 24 & 24\\ \hline
E1 & \multirow{5}{*}{Encodec} & \multirow{5}{*}{Encodec\_24k} & 1.5 &24 \\ 
E2 & & & 3 & 24\\ 
E3 & & & 6 & 24\\ 
E4 & & & 12 & 24\\ 
E5 & & & 24 & 24\\ \hline
F1 & \multirow{4}{*}{Funcodec}& en\_libritts\_16k\_nq32ds320 & 16 & 16\\ 
F2 & & en\_libritts\_16k\_nq32ds640 & 8 & 16\\ 
F3 & & zh\_en\_16k\_nq32ds320 & 16 & 16\\ 
F4 & & zh\_en\_16k\_nq32ds640 & 8 & 16\\ \hline
N & Soundstream & Soundstream & 6 & 16\\ \hline
M1 & \multirow{3}{*}{MP3} & - & 6 & -\\
M2 & & - & 24 & -\\
M3 & & - & 192 & -\\ \hline
O1 & \multirow{2}{*}{Opus} & - & 6 & -\\
O2 & & - & 24 & -\\ \hline
A1 & \multirow{3}{*}{AAC} & - & 6 & -\\ 
A2 & & - & 24 & -\\
A3 & & - & 192 & -\\ \hline

\end{tabular}
\end{table}

\subsection{Codec models}
We carefully select six cutting-edge, high-fidelity neural codec models for our comparative analysis, as shown in Table~\ref{tab:codec-inform}.
\textbf{Encodec} \cite{defossez2022highfi} serves as our baseline. 
We use the 2, 4, 8, 16, and 32 layer settings to compare with other codecs under similar bit rates.  
Building upon Encodec, \textbf{AudioDec} \cite{wu2023audiodec} introduces a novel approach employing group convolution to accelerate and streamline operations. 
\textbf{AcademiCodec} \cite{yang2023hifi} uses group-residual vector quantization to decrease the amount of codebook usage while keeping comparable performance. We use the universal version of the model. 
\textbf{FunCodec} \cite{du2023funcodec} proposed a frequency domain codec, which achieves a comparable performance with lower computation and parameter complexity. 
Additionally, \textbf{SpeechTokenizer} \cite{zhang2023speechtokenizer} introduces a unified speech tokenizer tailored for speech LMs, integrating HuBERT units as semantic teachers in the first layer of RVQ. 
\textbf{Descript Audio Codec (DAC)} \cite{kumar2024high} leverages advanced Snake activation from BigVGAN \cite{lee2022bigvgan} and utilizes a novel complex STFT discriminator at multiple time scales to further enhance audio fidelity. We use the 16k and 24k models.
Lastly, \textbf{SoundStream}\cite{zeghidour2021soundstreamendtoendneuralaudio}  use of RVQ in its encoders allows for a more efficient and compact representation of the audio signal, resulting in higher quality reconstruction at a lower bitrate.
\footnote{We use the implementation from: \url{https://github.com/kaiidams/soundstream-pytorch}}

To provide a more comprehensive comparison of the ability to retain emotional traits. we also chose three legacy codecs for comparisons: the \textbf{MP3} \cite{Brandenburg1994ISOMPEG1AA}, which is widely used for its efficient compression and good sound quality at different bitrates; the \textbf{Opus} \cite{valin2016highqualitylowdelaymusiccoding}, known for its adaptability to a wide range of audios and its low latency; and the \textbf{AAC} \cite{Bosi1997ISOIECMA}, which offers a high level of fidelity and is commonly used for streaming and broadcasting. These legacy codecs can help us benchmark neural codecs against established standards.

\begin{table}[t]
\centering
\fontsize{7}{9}\selectfont
\caption{The overview of the selected datasets and their setups. \textbf{Anno.} represents the annotation process we used. The \textbf{P} and \textbf{S} represent primary and secondary labeling scenarios, respectively.\label{tab:dataset-inform}}
\vspace{-0.3cm}
\begin{tabular}{c|c|c|c}
\toprule
\textbf{Datset} & \textbf{Language} & \textbf{Setting} & \textbf{Anno.}\\ 
\midrule
CREAM-D & English & Acted (Scripted) & P \\ 
IEMOCAP & English & Acted (Improvised \& Scripted) & S\\
IMPROV &  English & Acted (Improvised) & S\\ 
MSP-PODCAST & English & Real-world & P\\ 
BIIC-PODCAST & Chinese & Real-world & P\\ 
NNIME & Chinese & Acted (Improvised) & S\\ 
\bottomrule
\end{tabular}
\vspace{-0.3cm}
\end{table}
\subsection{Speech Emotion Recognition Dataset}
To assess the codecs, we use six public datasets partitioned by EMO-SUPERB \cite{wu2024emo}. Those datasets are classified according to their source (acted or real-world) and language (Chinese or English) shown in Table~\ref{tab:dataset-inform}. Datasets have two annotation scenarios: \textit{Primary} (P) asks each annotator to choose only one emotion during annotation, while \textit{Secondary} (S) allows annotators to choose multiple emotions for a clip. 

\section{Experimental Setup}
In contrast to prior SER approaches that relied on single target prediction \cite{Lee_2011,Chou_2019}, we employ a distribution-like representation to model the multi-dimensional complexity of emotion, as suggested by \cite{Chou_2024}. 
Furthermore, to enhance the performance of our SER model, we incorporate label smoothing \cite{Szegedy_2016} into the emotion distribution, effectively regularizing the classifier layer with a smoothing parameter set to 0.05.
\begin{figure*}[!t]
    \centering
    \vspace{-3cm}
    \includegraphics[width=\textwidth]{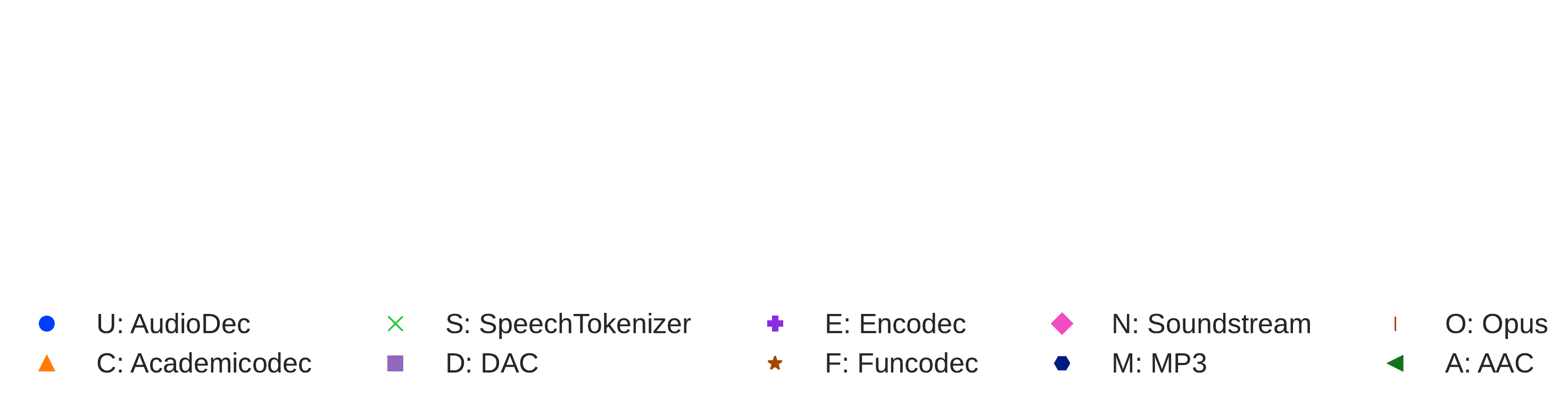}
    \vspace{-0.5cm}
    \caption{Legends for all codecs throughout this paper}
    \label{fig:codec_legend}
\end{figure*}

\subsection{Speech Emotion Recognition Models}
The model architecture is the same as the S3PRL \cite{yang21c_interspeech} toolkit, consisting of a CNN-Self Attention network. It includes three Conv1d layers, followed by a self-attention pooling layer, and two linear layers. We use class-balanced cross-entropy \cite{Cui_2019} to mitigate the effect of imbalance labels in emotion. 
We use a fixed learning rate of $10^{-4}$ and AdamW optimizer \cite{Loshchilov_2019} to train SER models until the loss of the development set stops decreasing for 5 epochs.

\subsection{Evaluation Metrics}
We use the macro-$F_1$ score \cite{opitz2019macro} as the evaluation metric for SER models. Since the output of SER models are probability distributions, an emotion prediction is successful if its probability surpasses a threshold $\frac{1}{n}$ for $n$-class SER models, following \cite{Riera_2019, Chou_2024}.

\subsection{Human Subjective Evaluation}

We are concerned with how human emotional perception of the synthesized sound differs from the original sound, so we conduct human subjective listening tests. We chose the audio samples from the well-known emotion database, the IEMOCAP. We randomly selected 45 pieces of audio and used various codecs to resynthesize the audio to conduct our subjective evaluation. We selected three codecs—Encodec, DAC, and the legacy codec Opus—using 6k and 24k sample rates. The annotators of the IEMOCAP dataset originally saw both video and audio to label emotions, but in our study, they only listened to the audio. This difference might affect emotion perception. Therefore, we also asked all annotators to label the emotions of the original audio for a fair comparison.

We hire annotators from the Prolific platform to annotate the resynthesized audio. Each audio is annotated by 5 male and 5 female annotators. We require annotators from the US and have more than 90\% of the acceptance rate. Each annotator answers three questions for each audio: (1) Choose one or more emotion(s) of the speaker from the same pre-defined options as the IEMOCAP. (2) What is the quality of the speech for the purposes of everyday speech communication on a scale of 1 to 5? (3) What is the quality of the speech based on the level of distortion of the speech on a scale of 1 to 5? 

After collecting annotations, we reported averaged values for speech quality and distortion factors. In terms of emotion, we calculate the distributional label for each sample, and we use the same way to use the threshold to binary the label for measuring the macro-$F_1$ score. We take the label of the original audio as ground truth and calculate the accuracy, which is defined \emph{subjective emotion loss}. 


\section{Result and Discussion}

\subsection{Effect of Bitrate on Human/Machine Emotion Recognition}
\begin{figure}[!t]
    \centering
    {\includegraphics[width=.48\textwidth]{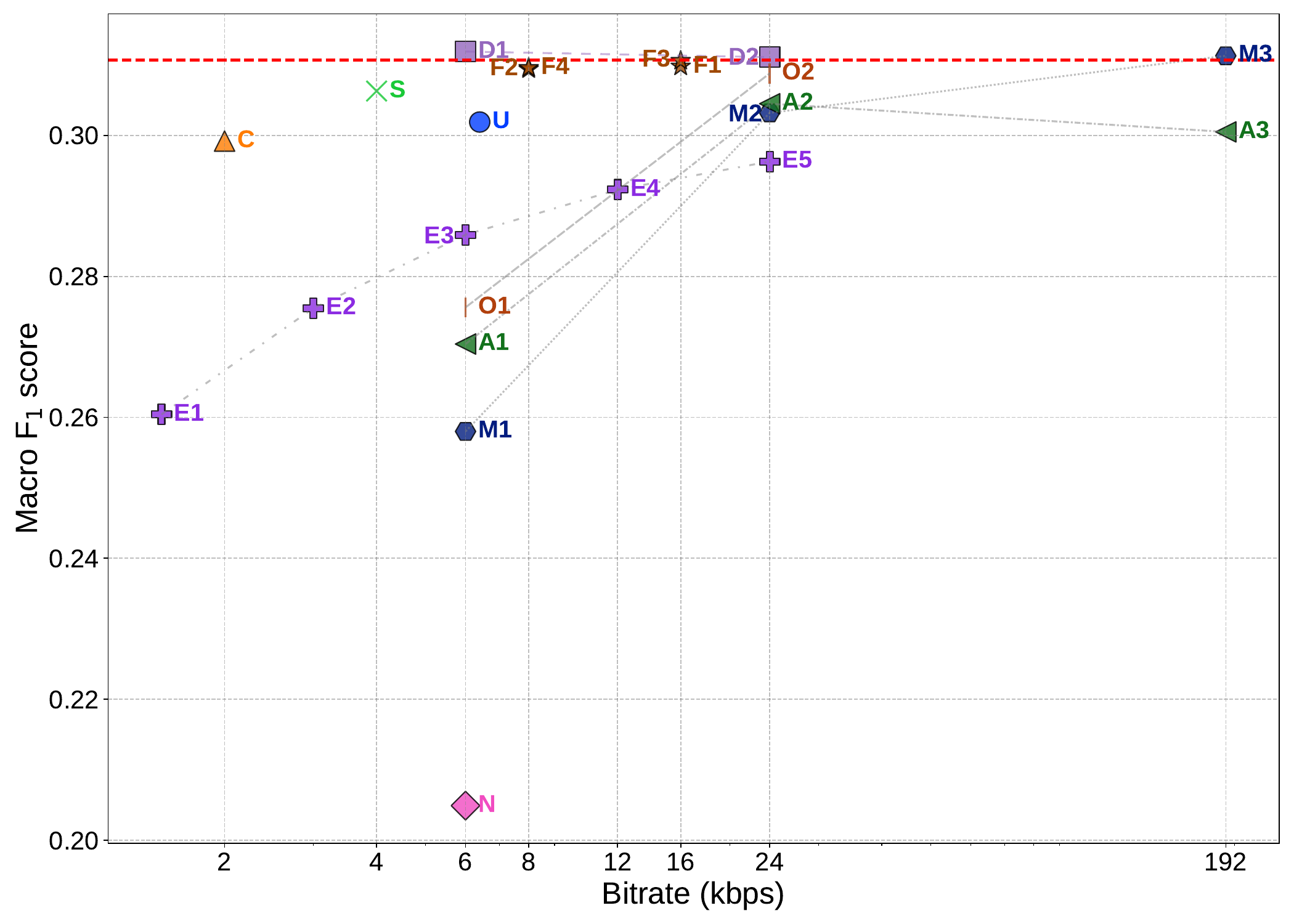}}
    \caption{Emotion recognition performance in macro-$F_1$ score on different codecs with the IEMOCAP dataset. The red dashed line represents the SER performance of the original audio.}
    \label{fig:IEMOCAP}
\end{figure}
Fig.~\ref{fig:IEMOCAP} shows the emotion retention capacity for codecs at different bitrates on the IEMOCAP dataset. In most codecs, both neural and legacy, there is a consistent trend: the emotion retention capacity is significantly enhanced as the bitrate of the codec increases. It highlights the direct impact of bitrate on the amount of voice information that can be transmitted and retained. A higher bitrate facilitates the retention of more detailed emotional information.

While lower bitrates usually impair the retention of emotional information, particular codecs have shown a remarkable emotion retention capacity even at low bitrates, such as the \textbf{SpeechTokenizer} and \textbf{Academicodec} series. This indicates that these codecs are particularly effective at maintaining emotional information's integrity despite lower bitrates' limitations. Among all codecs, Soundstream was not good at retaining emotional information, showing the worst emotion recognition performance than all the other codecs.

When comparing legacy codecs (represented by \textbf{MP3}, \textbf{Opus}, \textbf{AAC}) with neural codecs, legacy codecs typically show significant improvements in emotional information retention at higher bitrates, e.g. the \textbf{Opus} series codecs perform well at higher bitrates, approaching baseline performance, the \textbf{MP3} series codecs show varying performance, with some models (e.g. MP3\_24k) maintaining good emotional information retention even at higher bitrates. In contrast, others (e.g. MP3\_6k) do not perform well at low bitrates, with the \textbf{AAC} series of legacy codecs in a similar situation.

In contrast, the neural codecs in the DAC family consistently outperform legacy codecs at the same bit rate, highlighting their advanced design and efficiency in retaining emotional information. DAC's superior performance might be attributed to two tricks: snake activation function, and balanced data sampling during training. It also performs well at 6kbps, suggesting that neural codecs are better suited to retaining the integrity of emotional information in bitrate-constrained environments.


\subsection{Trend Across English Datasets}

\begin{figure*}[!t]
    \centering
    \captionsetup[subfigure]{position=top}
    \subfloat[CREMA-D]{\includegraphics[width=.33\textwidth]{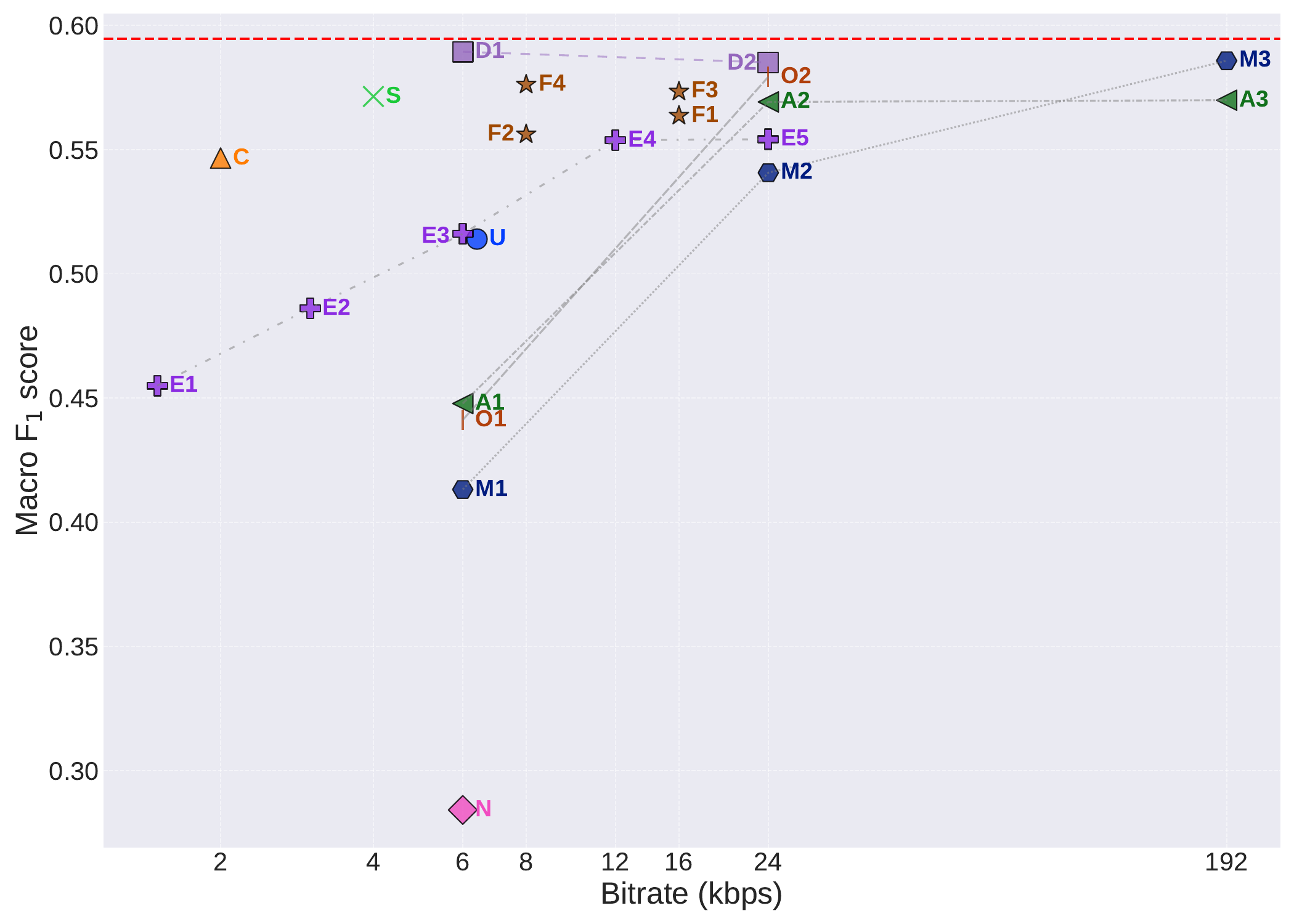}}\subfloat[IMPROV]{\includegraphics[width=.33\textwidth]{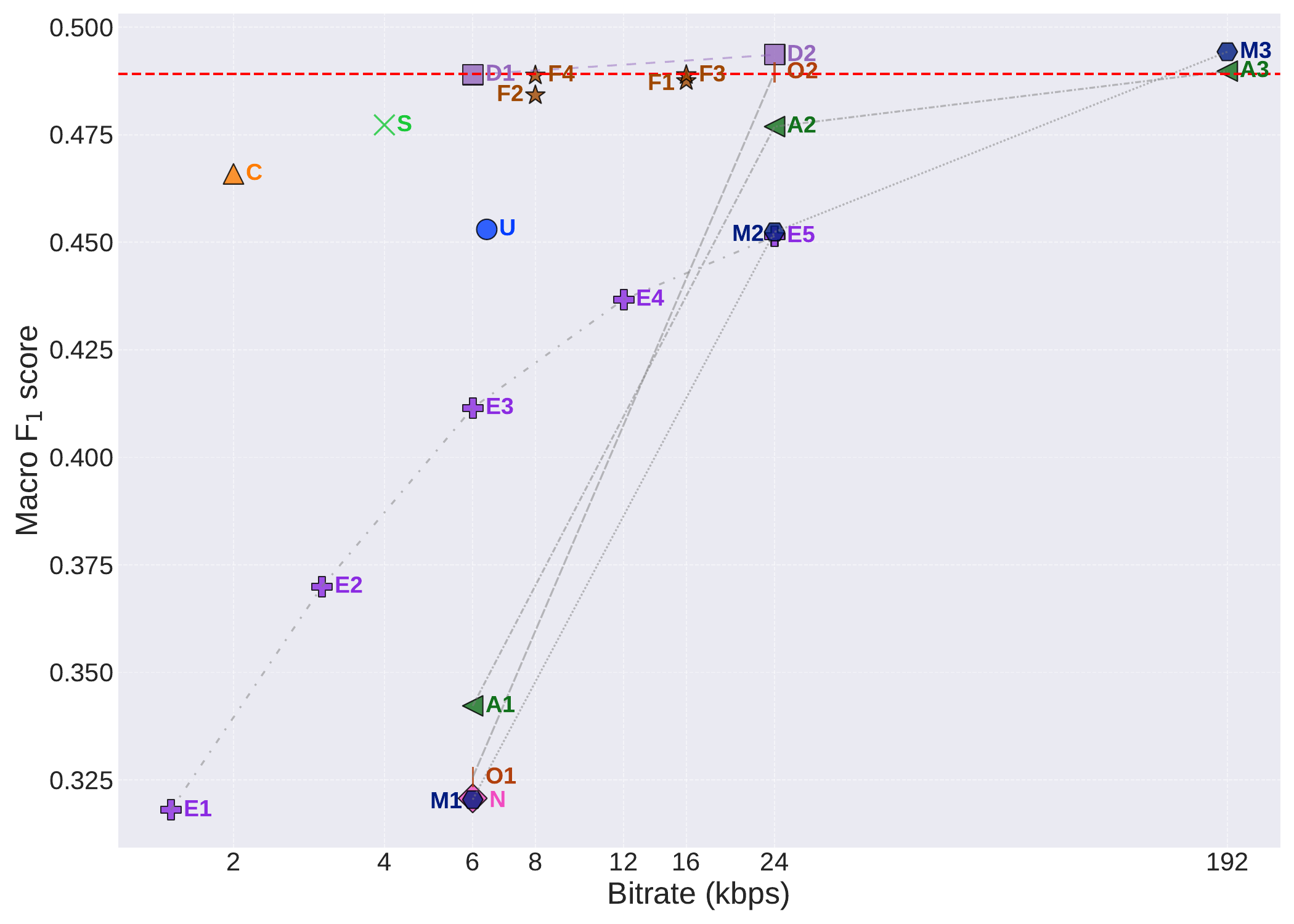}}\subfloat[PODCAST]{\includegraphics[width=.33\textwidth]{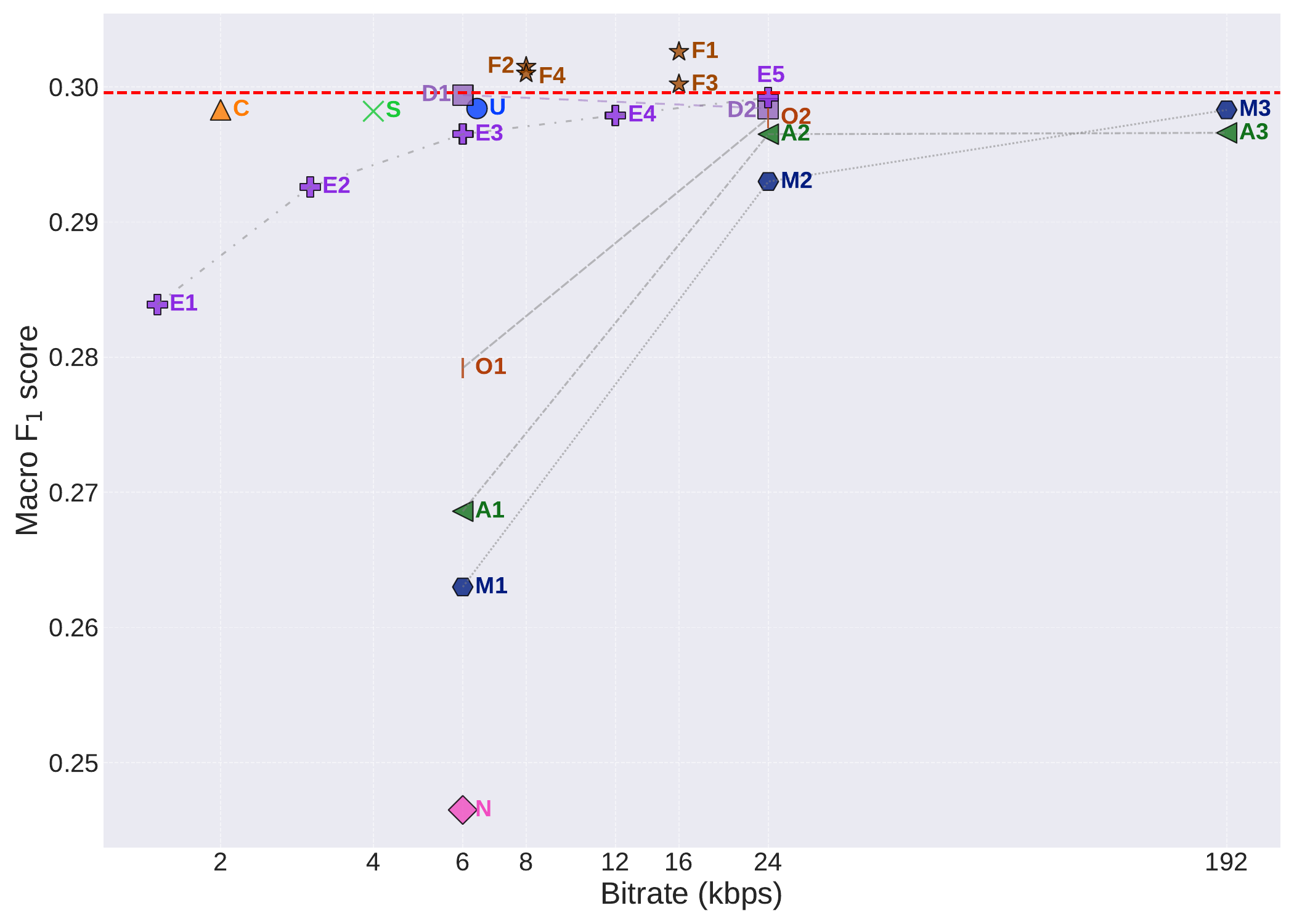}}
    \caption{Emotion recognition performance (macro-$F_1$) on different codec with (a) CREMA-D (b) IMPORV (c) PODCAST dataset. The red dashed line represents the SER performance of the original audio.}
    \label{fig:three_english_dataset}
\end{figure*}

This section discusses the emotional information retention capacity of codec models in three additional English datasets (CREMA-D, IMPROV, and PODCAST). This analysis extends the performance of codecs in the IEMOCAP dataset from Section~\ref{fig:IEMOCAP}, highlighting consistent trends and codec efficacy on emotional information retention capacity.

According to Fig. 4, higher bitrates generally improve the retention of emotional information in English datasets. Neural codecs, particularly those in the \textbf{DAC} family, excel at preserving emotions even at lower bitrates (6 kbps), outperforming other codecs. The \textbf{Funcodec} series matches baseline performance and achieves the highest score on the PODCAST dataset. Neural codecs like \textbf{SpeechTokenizer} and \textbf{Academicdec} also perform well at low bitrates, demonstrating their advanced design and efficiency in retaining emotional nuances. In contrast, legacy codecs such as \textbf{Opus} and \textbf{MP3} show improvement at higher bitrates but still lag behind neural codecs in preserving emotional information. Overall, codecs with higher bitrates generally enhance emotional capacity, with neural codecs consistently outperforming legacy ones across all bitrates.



Legacy codecs fall short compared to the top-performing neural codecs, highlighting the importance of neural codecs for retaining emotional information in bitrate-constrained environments.

\subsection{Variability Across Chinese Datasets}

\begin{figure*}[!t]
    \centering
    \captionsetup[subfigure]{position=top}
    \subfloat[NNIME]{\includegraphics[width=.4\textwidth]{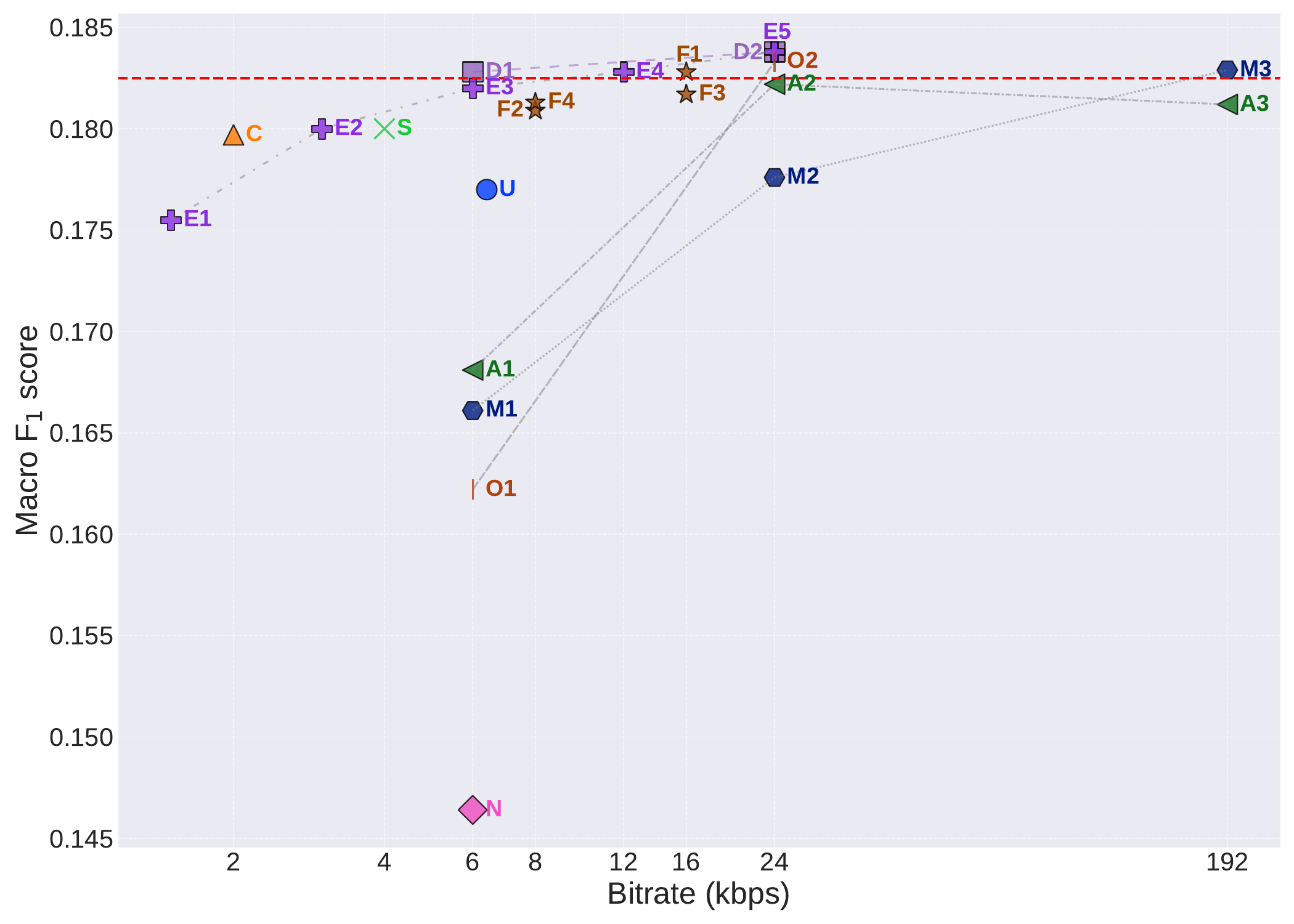}}\subfloat[BIIC-PODCAST]{\includegraphics[width=.4\textwidth]{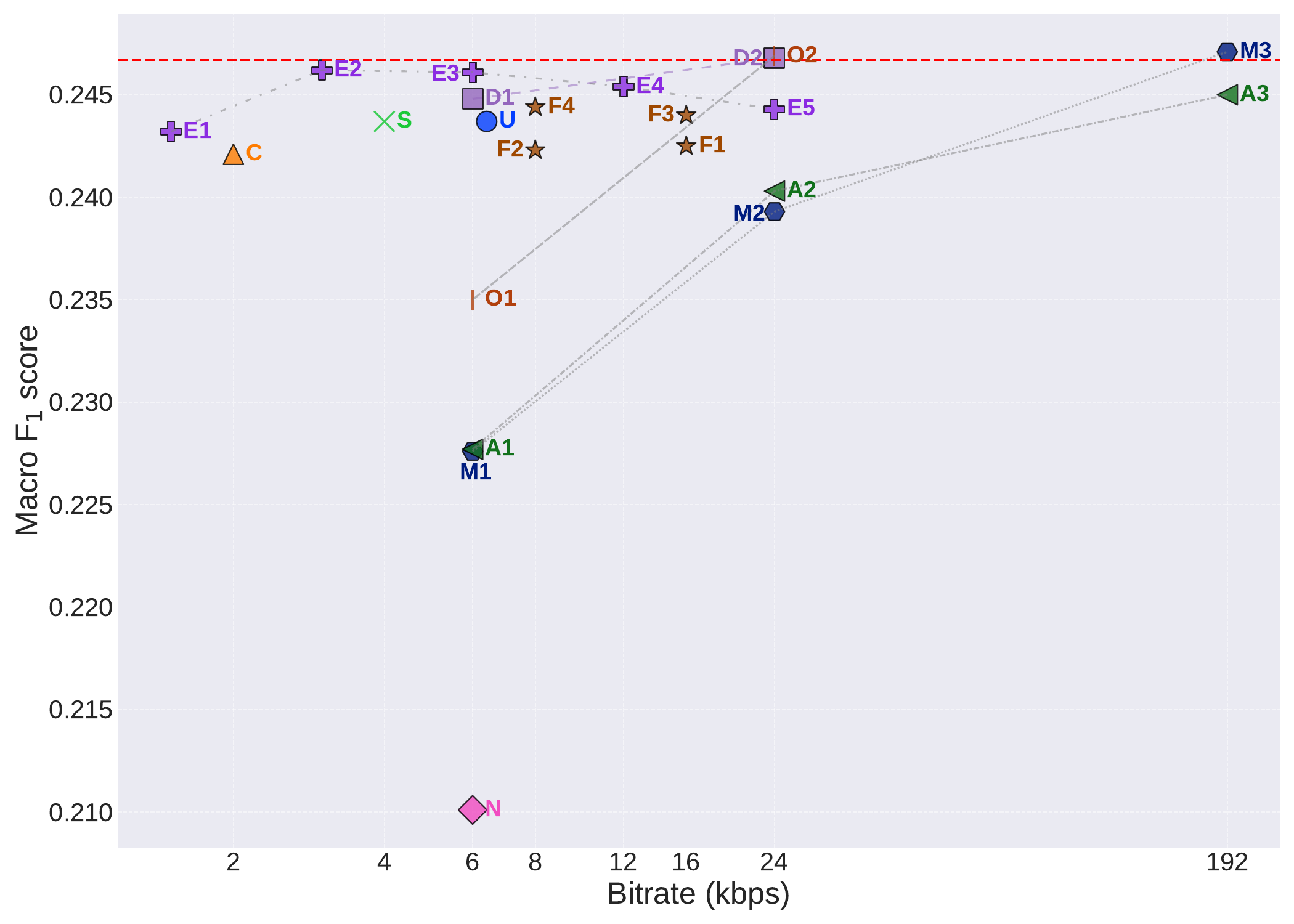}}
    \caption{Emotion recognition performance (macro-$F_1$) on different codec with (a) NNIME (b) BIIC-PODCAST Chinese datasets. The red dashed line represents the SER performance of the original audio.}
    \label{fig:two_chinese_dataset}
\end{figure*}


Fig.~\ref{fig:two_chinese_dataset} illustrate the emotion preservation capacity of various codecs at different bitrates on Chinese datasets BIIC-PODCAST and NNIME. 

Similar to the English Dataset, higher bitrate codecs typically retain emotional information better in the Chinese dataset; the \textbf{DAC} series still delivers excellent performance compared to codecs at the same bitrate; neural codecs like \textbf{SpeechTokenizer} and \textbf{Academicodec} also maintain good performance at lower bitrate; legacy codecs were consistently weaker than neural codecs.

While the overall trend is consistent, there are some differences when comparing in detail the performance of some codecs on English and Chinese datasets.

In the case of the \textbf{Encodec} series, a significant increase in emotional information retention with increasing bitrate can be observed in the English dataset. However, in the Chinese dataset, this trend slows down as the bitrate increases and even decreases in the BIIC-PODCAST dataset. Compared with English, the emotional information in Chinese have different phonetic and tonal characteristics, and the neural codec may not have been specifically optimised for these differences, resulting in a slow increase or even a decrease in the efficiency of emotional information retention with increasing bitrate.

\begin{figure}[!t]
    \centering
    {\includegraphics[width=.5\textwidth]{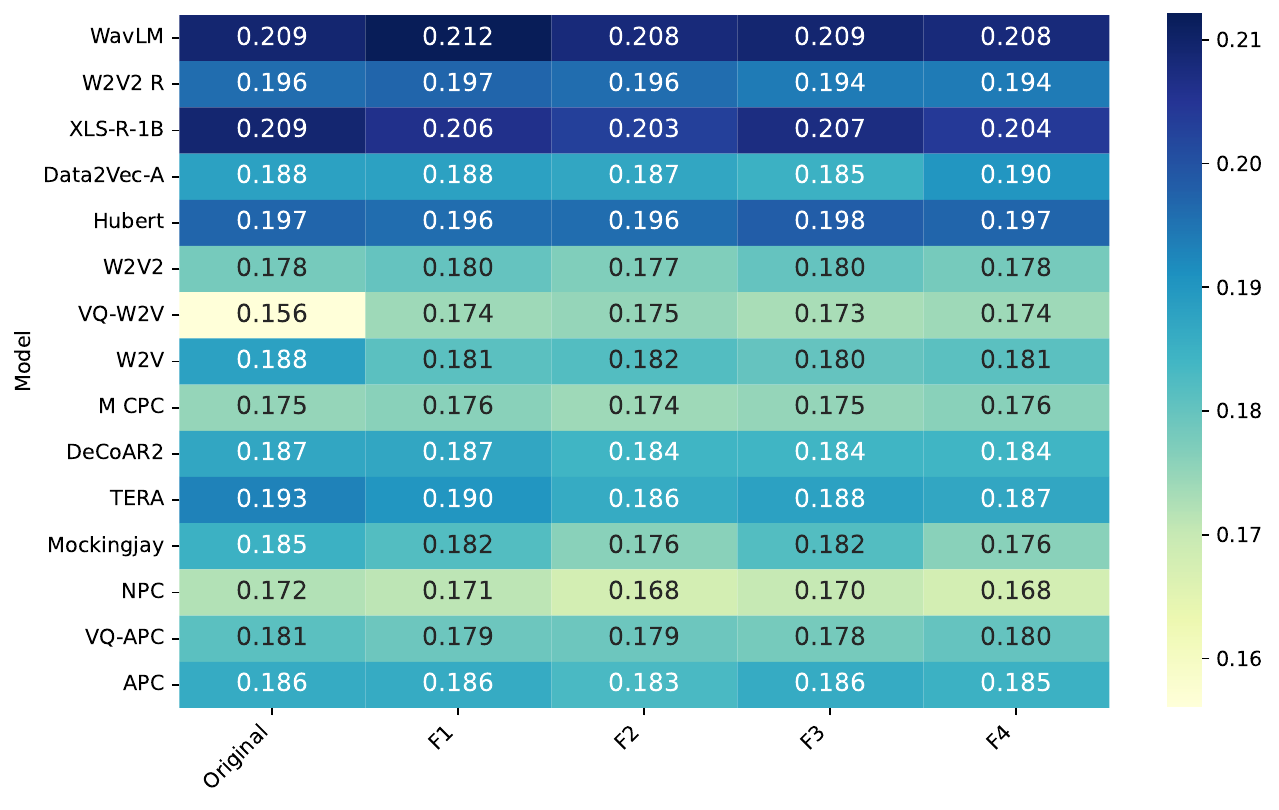}}
    \caption{SER macro-$F_1$ scores of resynthesized emotional
speech from the NNIME dataset on FunCodec models.}
    \label{fig:Funcodec_results}
\end{figure}

Also, specific codecs are trained on English-only datasets, such as \textbf{funcodec\_en\_libritts}, while on the contrary,  \textbf{funcodec\_zh\_en}, trained on a mixture of English and Chinese datasets. The insight received from Fig.~\ref{fig:Funcodec_results} is that the neural codec model trained on the Chinese dataset BIIC-PODCAST dataset with a mixture of English and Chinese is slightly better at preserving Chinese emotional information than the neural codec model trained on the English-only dataset. However, the \textbf{funcodec\_zh\_en} series did not beat the \textbf{funcodec\_en\_libritts} series on the Chinese dataset NNIME. Such an observation motivates the study of more codec model-trained datasets to accommodate the generic representation of different emotional information across languages.


\subsection{Specific Emotion Losses}
\begin{table*}[!t]
\centering
\fontsize{8}{9}\selectfont
\caption{SER $F_1$ score drop of resynthesized emotional speech on different emotions relative to $F_1$ score of the original speech, in \%. Positive values imply an increase in SER performance. Emotions include depression (P), frustration (T), anger (A) sadness (S), disgust (D), excitement (E), fear (F), neutral (N), surprise (U), and happiness (H). The row \textbf{Ori.}, represents the performance of original audio in $F_1$ score.}
\vspace{-0.3cm}
\setlength{\tabcolsep}{2pt}
\renewcommand{\arraystretch}{0.86}
\begin{tabular}{c|cccccccccc|ccccccccc}
\toprule
 & \multicolumn{10}{c|}{IMPROVS} & \multicolumn{9}{c}{IEMOCAP}\\
\midrule
 ID & P & T & A & S & D & E & F & N & U & H & T & A & S & D & E & F & N & S & H\\
\midrule
Ori. & 38.16 & 53.34 & 47.91 & 57.33 & 14.35 & 57.07 & 1.86 & 84.04 & 29.12 & 69.21 & 69.27 & 59.70 & 61.36 & 00.89 & 55.10 & 9.58 & 67.54 & 25.19 & 49.28\\ \midrule
U & -50.49 & -11.88 & -8.68 & -19.15 & -35.13 & 2.05 & -70.65 & -0.51 & -6.90 & -3.37 & -1.93 & -3.06 & -5.62 & -30.80 & -4.20 & -7.76 & -2.67 & -12.10 & -6.99\\
C & -42.27 & -8.22 & -5.96 & -13.03 & -25.01 & 1.37 & -73.60 & -0.48 & 4.85 & -2.45 & -1.81 & -4.79 & -8.90 & -25.63 & -5.03 & -20.83 & -6.77 & -8.13 & -7.44\\
S & -23.12 & -5.39 & -5.61 & -5.93 & -24.68 & 0.36 & -48.77 & -0.31 & -3.41 & -1.88 & -1.37 & -1.67 & -4.07 & -3.81 & -2.62 & -11.04 & -1.78 & -7.79 & -4.36\\
D1 & -7.58 & -1.75 & -1.37 & -2.06 & -3.44 & 0.64 & -23.91 & -0.16 & -1.76 & -0.47 & -0.31 & -0.19 & -0.72 & -8.84 & -0.47 & -2.97 & -0.40 & 0.22 & -0.72\\
D2 & -1.84 & -0.80 & -1.21 & -0.37 & -9.82 & 0.65 & -21.38 & -0.08 & 1.28 & -0.14 & -0.02 & -0.40 & -0.40 & -27.97 & -0.31 & -1.81 & -0.13 & 0.16 & -0.30\\
E1 & -88.21 & -42.40 & -22.08 & -68.06 & -70.39 & -7.00 & -94.65 & -16.25 & -28.31 & -6.26 & -4.91 & -10.47 & -37.24 & -19.09 & -12.82 & -41.13 & -28.06 & -34.14 & -18.71\\
E5 & -43.75 & -6.24 & -2.94 & -19.94 & -39.15 & 1.84 & -57.34 & -1.22 & -3.94 & -3.39 & -1.43 & -3.90 & -12.28 & -10.62 & -5.65 & -18.05 & -6.68 & -10.58 & -5.62\\
F1 & -12.57 & -0.54 & -1.63 & -3.10 & -10.31 & 1.15 & -26.03 & -0.07 & -0.07 & -0.82 & -0.72 & -0.47 & -1.12 & -5.97 & -1.73 & -4.23 & -0.55 & -5.07 & -1.48\\
F3 & -11.86 & -1.29 & -2.03 & -3.25 & -11.97 & 1.08 & -30.95 & -0.10 & 3.63 & -0.61 & -1.02 & -0.98 & -1.34 & -24.16 & -1.78 & -9.88 & -1.08 & -1.00 & -1.70\\
M1 & -76.03 & -62.86 & -48.19 & -57.12 & -88.15 & -14.26 & -94.21 & -15.82 & -4.42 & -8.45 & -16.58	& -20.97 & -23.86 &	1.29 & -17.02 &	-54.93	& -14.26 & -33.99 &	-28.78\\
M3 & -8.07 & -5.33 & -7.00 & -2.82 & -1.36 & -11.41 & -40.98 & 0.49 & -9.42 & -2.08 & -0.21 & -0.10 & -3.99 & -28.14 & -1.54 & 0.28 & -0.63 & -5.75 & -5.41\\
O1 & -87.75 & -44.84 & -27.26 & -68.92 & -81.77 & -12.65 & -92.32 & -15.93 & -35.08 & -7.74 & -10.67 & -14.65 & -19.91 & -19.66 & -14.21 & -37.22 & -7.63 & -33.98 & -21.71\\
O2 & -23.29 & -5.32 & -6.04 & -6.27 & -5.44 & -6.98 & -49.64 & 0.51 & -11.87 & -2.24 & -0.28 & -1.18 & -4.34 & -18.19 & -1.82 & -6.20 & -1.46 & -7.46 & -7.00\\
A1 & -57.40	& -61.04 & -52.67 & -38.45 & -94.45 & -11.11 & -95.81 & -14.95 & -15.95 & -6.56 & -10.71 & -15.80 & -19.79 & -12.93 & -17.42 & -49.14 & -11.81 & -30.28 & -25.50\\
A3 & -15.53 & -6.24	& -8.17	& -5.52	& -8.68	& -10.30 & -38.00 & 0.21	& -19.70 & -1.83 & -1.29 & -3.92 & -10.34 & -46.57 & -3.02 & -11.84 & -2.71 & -17.44 & -9.60\\ \midrule
Avg & -36.65 & -17.61 & -13.39 & -20.93 & -33.98 & -4.30 & -57.22 & -4.31 & -8.74 & -3.22 & -3.55 & -5.50 & -10.26 & -18.74 & -5.98 & -18.45 & -5.77 & -13.82 & -9.69
\\ \bottomrule
\end{tabular}
\label{tab:emo_drop}
\end{table*}


From Table 4, speech resynthesized through a codec may significantly lose certain emotional information, particularly for emotions challenging for speech emotion recognition (SER) models, such as disgust or fear. For instance, the $F_1$ scores for the fear emotion category on the IMPROV and IEMOCAP datasets are only 1.86\% and 9.58\%, respectively. The $F_1$ scores for the resynthesized speech for fear dropped by 51.87\% and 15.03\%, significantly higher than the other emotions. This indicates that when the emotion recognition performance on the original audio is low, any additional distortion from the codec has a pronounced impact. Despite the low absolute scores, the considerable percentage drop underscores the codec's inability to retain emotional information of fear, which is already a challenge for the model's detection.
In addition to fear, other emotions such as depression and sadness also experience degraded SER performance when speech is resynthesized through these codecs. This highlights the shortcomings of codecs in retaining complex and subtle emotional information. While codecs are proficient in compressing and resynthesizing speech, they struggle to preserve the nuanced emotional cues necessary for accurate emotion recognition, particularly for more complex emotions. This demonstrates the necessity for further optimization and enhancement of codec models to maintain the emotional integrity of resynthesized speech better.

\begin{figure}[!t]
    \centering
    \begin{minipage}{0.45\textwidth}
        \centering
        \subfloat[Communication]{%
            \includegraphics[width=0.48\textwidth]{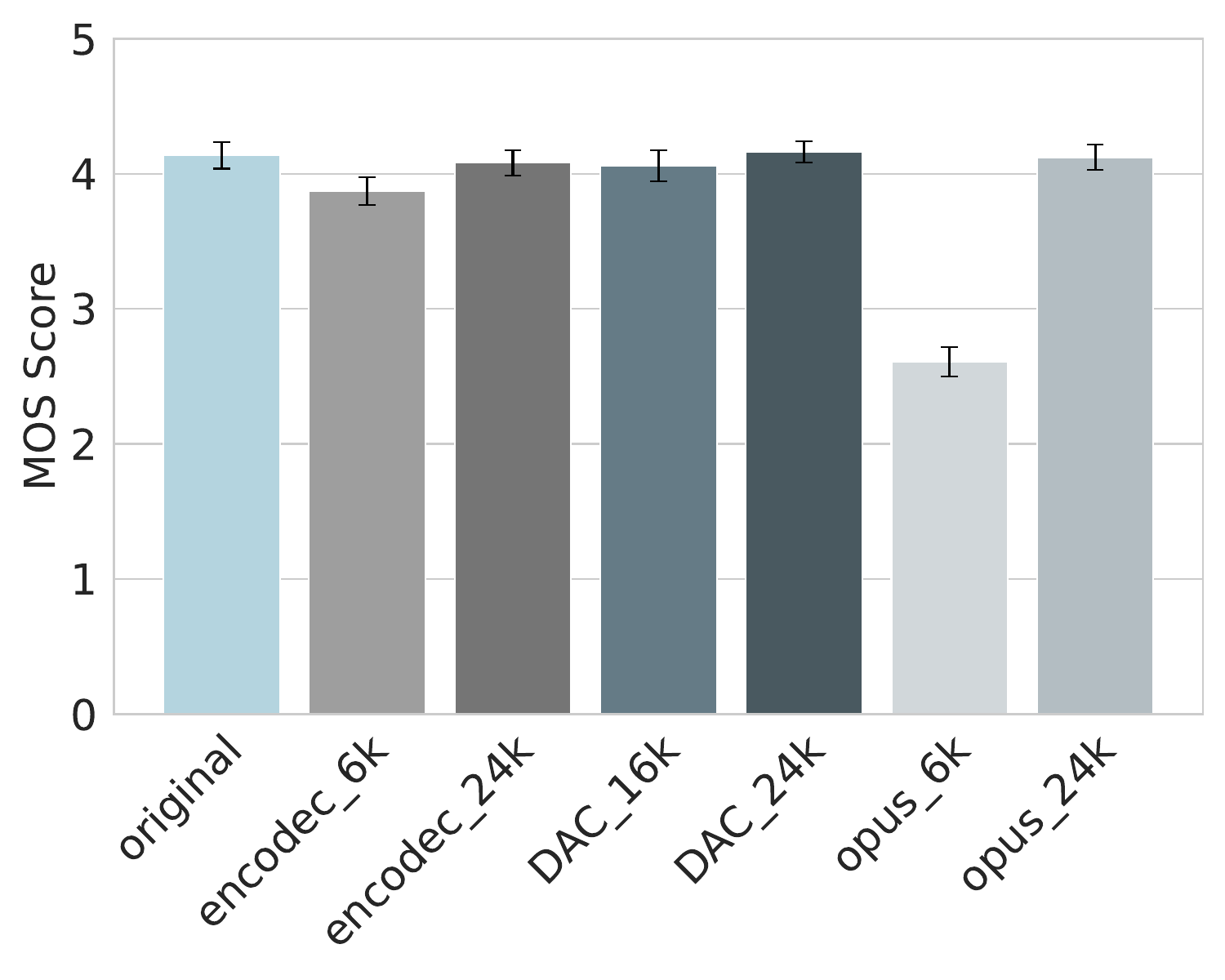}
        }
        \hfill
        \subfloat[Distortion]{%
            \includegraphics[width=0.48\textwidth]{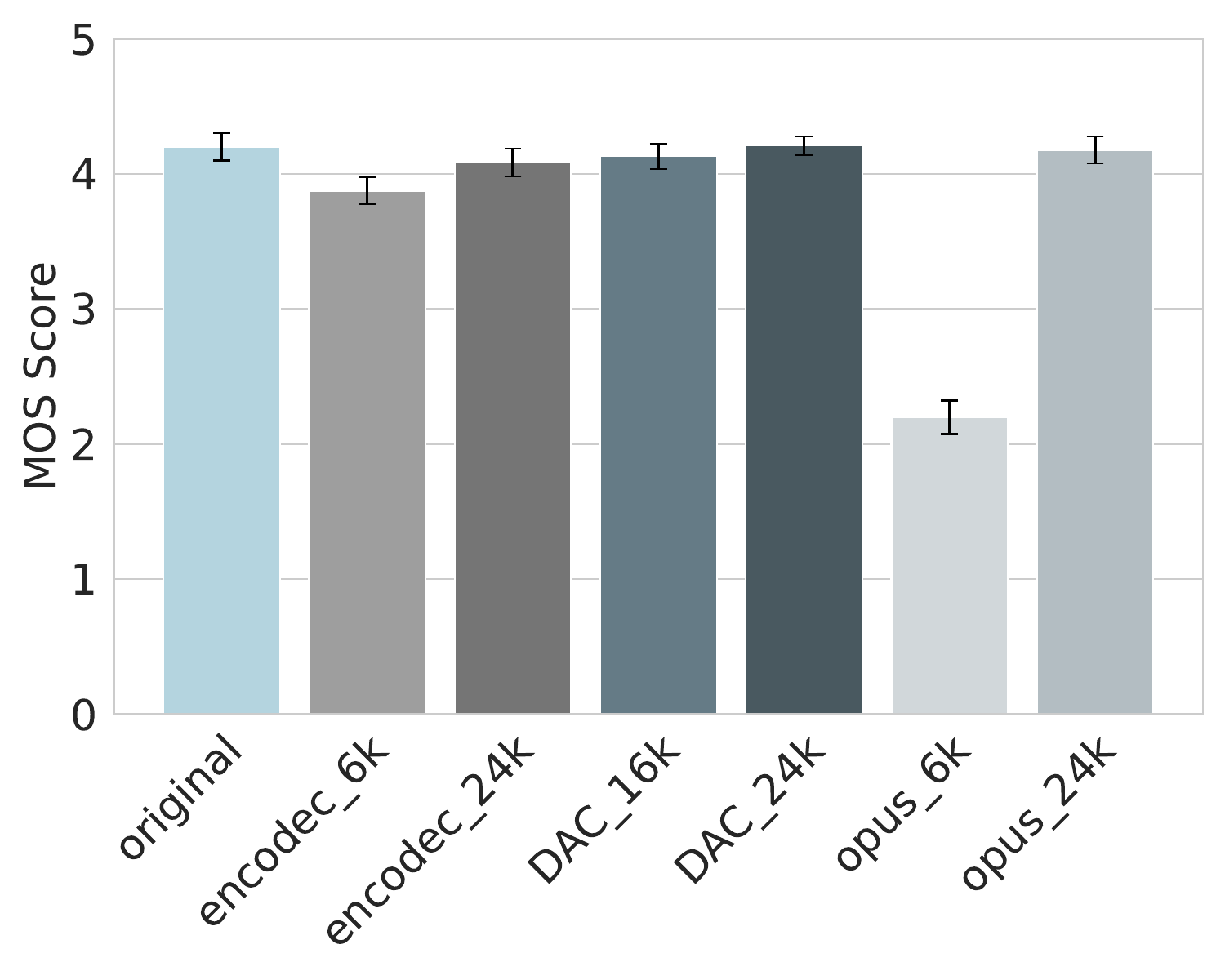}
        }
    \end{minipage}
    \hfill
    \begin{minipage}{0.45\textwidth}
    \end{minipage}
    \caption{Speech subjective quality evaluation (MOS score) based on (a) everyday speech communication and (b) distortion, following ITUT P835 \cite{itut}. A higher score indicates a higher quality.}
    \label{fig:quality}
\end{figure}

\begin{figure}[!t]
    \centering
    {\includegraphics[width=.45\linewidth]{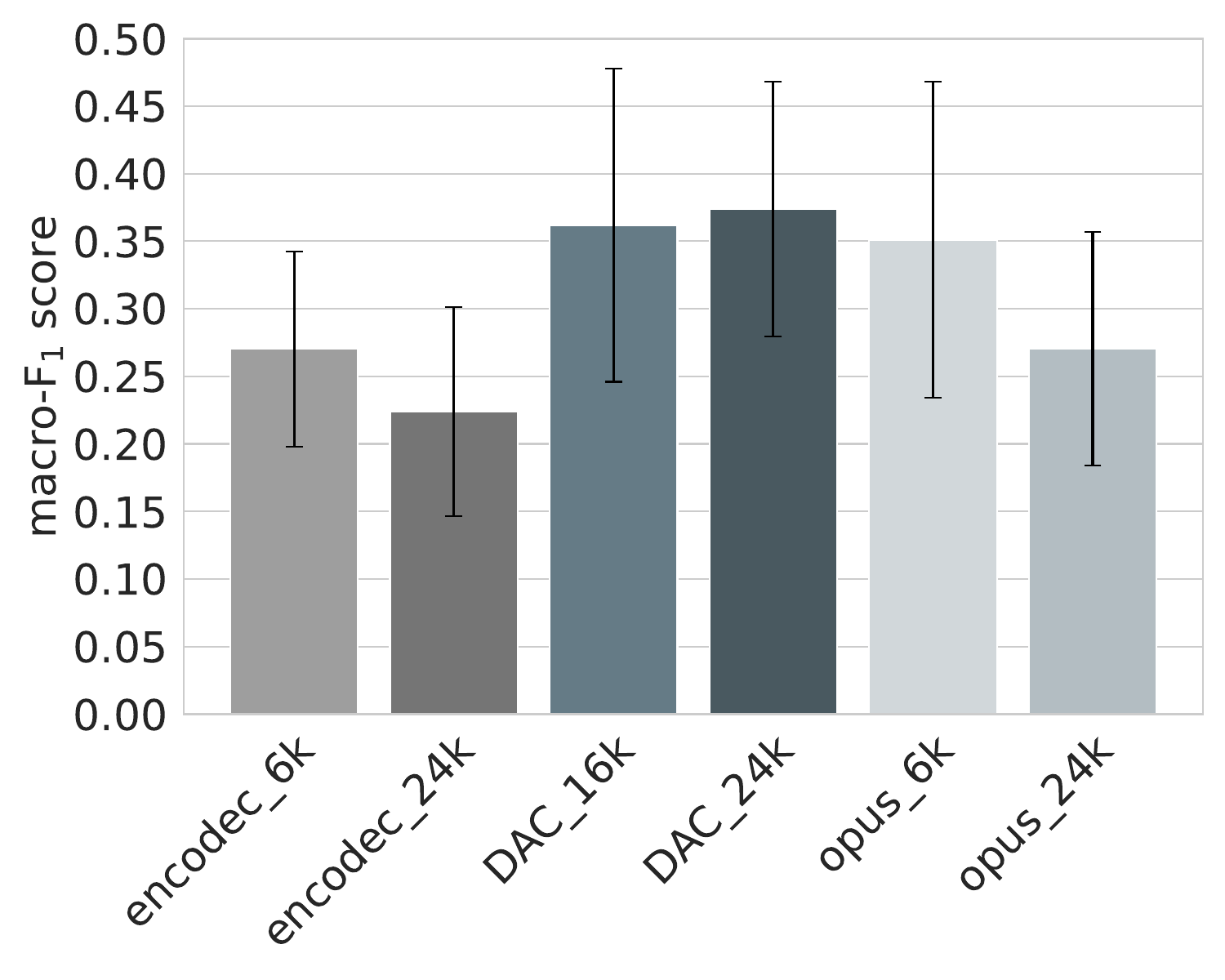}}
    \caption{Subjective emotion retention capacity (macro-$F_1$ score) of six codecs examined.}
    \label{fig:hum_sub_results}
\end{figure}

\subsection{Human Subjective Listening Tests}
Fig.~\ref{fig:quality} shows the Mean Opinion Score (MOS) of original audio and audio resynthesized by codecs. The original audio has the highest MOS. Codecs with a 24k bitrate experience a slight drop in MOS but still maintain scores close to the original audio. In contrast, codecs with a 6k bitrate show a significant drop in MOS, except for DAC 16k, which maintains a high MOS. Opus 6k has the lowest MOS among our evaluations. These observations align with our findings on the objective emotion preservation capacity of these codecs.

Fig.~\ref{fig:hum_sub_results} illustrates macro-$F_1$ scores calculated by the emotional ratings from workers on the original audios and resynthesized audios. We assume the ground truth is the labels of the original audio, and the resynthesized ones are predicted labels. We define the score as \textbf{subjective\ emotion retention capacity}. Based on the results, the \textbf{DAC 24K} has the highest emotion retention capacity, but the \textbf{Encodec 24K} has the lowest performance.  The findings align with the objective evaluation in Fig.~\ref{fig:IEMOCAP}.

\section{Limitation}
Existing neural codec models are predominantly trained by English and Chinese datasets. However, there are thousands of spoken languages around the world. Whether existing codec models can be generalized to other languages and how to train codecs that preserve multilingual emotion information are issues that have not been solved. 

While subjective human evaluations provided valuable insights, the pool of annotators was limited to a specific demographic. Future studies should aim to include a more diverse group of evaluators to account for varying perceptions of emotions across different age groups, cultures, and backgrounds.

Due to limited scope, this work does not consider how the interactivity and context of a conversation might influence the effectiveness of emotion preservation. Future research should explore how codecs perform in interactive dialogue systems where context and conversational history play crucial roles in emotion recognition and response generation.

\section{CONCLUSION}
This work provides extensive insights into the capacity of emotional information preservation of neural codec models.
We provide a different perspective to evaluate the performance of codecs.
We find that codecs with higher bitrate preserve more emotional information. 
DAC performs best among all neural codecs, while AcademiCodec and SpeechTokenizer preserve considerable emotional information under a limited bit rate, and our subjective evaluation aligns with the findings.
Legacy codecs perform worse than neural codecs at low-bitrate scenarios.
Furthermore, training codecs with Chinese and English data might have a limited effect on Chinese Emotion recognition compared with codecs trained with English data. 
We find that the resynthesis of speech by neural codecs degrades the information in emotions such as sadness, depression, fear, and disgust.
Our future work will train codec models to preserve emotional information across diverse linguistic contexts. 

\bibliographystyle{IEEEbib}
\bibliography{strings,refs}
\clearpage

\appendix

\setcounter{table}{0}
\setcounter{figure}{0}

\end{document}